# Far infrared measurements of absorptions by $CH_4+CO_2$ and $H_2+CO_2$ mixtures and implications for greenhouse warming on early Mars

Martin Turbet[a], Ha Tran[a], Olivier Pirali[b,c], François Forget[a],
Christian Boulet[b], Jean-Michel Hartmann[d,#]

[a] Laboratoire de Météorologie Dynamique/IPSL, CNRS, Sorbonne Université, Ecole normale supérieure, PSL Research University, Ecole Polytechnique, 75005 Paris, France.
[b] Institut des Sciences Moléculaires d'Orsay (ISMO), CNRS, Université Paris-Sud, Université Paris-Saclay, F-91405 Orsay, France.
[c] AILES beamline SOLEIL Synchrotron, L'Orme des Merisiers, 91190 Saint-Aubin, France
[d] Laboratoire de Météorologie Dynamique/IPSL, CNRS, Ecole polytechnique, Sorbonne Université, Ecole normale supérieure, PSL Research University, 91128 Palaiseau, France.
[#] Corresponding author: jean-michel.hartmann@lmd.polytechnique.fr

**Abstract**

We present an experimental study of the absorption, between 40 and 640 $cm^{-1}$, by $CO_2$, $CH_4$ and $H_2$ gases as well as by $H_2+CO_2$ and $CH_4+CO_2$ mixtures at room temperature. A Fourier transform spectrometer associated to a multi-pass cell, whose optics were adjusted to obtain a 152m path length, were used to record transmission spectra at total pressures up to about 0.98 bar. These measurements provide information concerning the collision-induced absorption (CIA) bands as well as about the wing of the $CO_2$ 15 µm band. Our results for the CIAs of pure gases are, within uncertainties, in agreement with previous determinations, validating our experimental and data analysis procedures. We then consider the CIAs by $H_2+CO_2$ and $CH_4+CO_2$ and the low frequency wing of the pure $CO_2$ 15 µm band, for which there are, to our knowledge, no previous measurements. We confirm experimentally the theoretical prediction of Wordsworth et al. 2017 that the $H_2+CO_2$ and $CH_4+CO_2$ CIAs are significantly stronger in the 50-550 $cm^{-1}$ region than those of $H_2+N_2$ and $CH_4+N_2$, respectively. However, we find that the shape and the strength of these recorded CIAs differ from the aforementioned predictions. For the pure $CO_2$ line-wings, we show that both the $\chi$-factor deduced from measurements near 4 µm and a line-mixing model very well describe the observed strongly sub-Lorentzian behavior in the 500-600 $cm^{-1}$ region. These experimental results open renewed perspectives for studies of the past climate of Mars and extrasolar analogues.

## 1. Introduction

It is well known that, in the atmospheres of many planets, the far wings (FWs) of molecular absorption lines (e.g. those of $CO_2$ in Venus, of $CH_4$ in Titan and Jupiter) and collision-induced absorptions (CIAs, e.g. that by $CO_2$ pairs in Venus; of $N_2$ pairs in Titan; of $H_2$ pairs in Jupiter) significantly contribute to the radiative budgets. This has motivated numerous experimental and theoretical laboratory studies of the CIA of various molecular systems, from the visible to the far infrared (see reviews in Frommhold 2006, Hartmann et al. 2008, Richard et al. 2012, Hartmann et al. 2018b and Karman et al. 2018). The same remark stands for the FWs, but the latter were essentially studied for $H_2O$ and $CO_2$ only (see Hartmann et al. 2008, Hartmann et al. 2018b and references therein).

Understanding how the early Martian climate could have been warm enough for liquid water to flow on the surface remains one of the major enigmas of planetary science (Wordsworth 2016, Haberle et al. 2017) and no consensus scenario has yet been reached. Thick, $CO_2$-dominated atmospheres do not provide the necessary greenhouse effect to warm the surface of early Mars above the melting point of water (Forget et al. 2013, Wordsworth et al. 2013). In this case, longwave radiation losses essentially occur in the 200-600 $cm^{-1}$ region (Wordsworth et al. 2010) where absorption and emission are dominated by the $CO_2$ roto-translational CIA and the FWs of the intense 15 µm-band lines. The radiative budget in this spectral region is significantly affected by the assumption made on the shape of these FWs (Turbet & Tran 2017), and no measurements are, to our knowledge, available to constrain it.

In reducing atmosphere conditions, absorptions can arise from the CIAs of $H_2$-$CO_2$ and $CH_4$-$CO_2$ pairs. Such absorptions can potentially absorb in the 200-600 $cm^{-1}$ region corresponding to a carbon dioxide transparency window and thus produce a strong greenhouse warming in a $CO_2$-dominated atmosphere. Up to now, the modelings of these two contributions were based, due to lack of relevant data, on the CIAs of $H_2$-$N_2$ and $CH_4$-$N_2$ pairs, respectively (e.g. Ramirez et al. 2014). Wordsworth et al. 2017 recently calculated that the collision-induced absorptions by $H_2$+$CO_2$ and $CH_4$+$CO_2$ mixtures should be significantly stronger than those for $H_2$+$N_2$ and $CH_4$+$N_2$, respectively. These calculations have important implications for our understanding of the climate of early Mars, at the time when valley networks were carved by liquid water. Indeed, with these new CIAs, only few % of $CH_4$ and/or $H_2$ in a $CO_2$-dominated atmosphere could suffice to warm early Mars enough for surface liquid water to become stable (Wordsworth et al. 2017, Ramirez 2017). These calculated CIAs, if confirmed by laboratory experiments, have the power to reconcile the Martian geology and mineralogy with our knowledge on atmospheric sciences (see Wordsworth et al. 2017 and references therein).

The remainder of this paper is organized as follows. Section 2 describes the experiments and the way the recorded spectra of $CO_2$, $CH_4$, $H_2$, $H_2$+$CO_2$ and $CH_4$+$CO_2$ were analyzed. The results obtained for the CIAs and for the low frequency wings of the pure $CO_2$ lines are presented in Sections 3 and 4, respectively. Discussions of the results and the consequences for studies of the past climate of Mars are the subject of Section 5. Final conclusions and possible directions for future studies are given in Section 6.

## 2. Experiments and their analysis

### 2.1   Spectra recordings

The spectra used in this study have been recorded at the AILES line of the Soleil synchrotron facility in a way similar to that used in Pirali et al. 2009 and Hartmann et al. 2011. A Bruker IFS 125HR Fourier transform spectrometer (FTS) and a 2.5 m long multi-pass cell with 60 µm thick polypropylene windows, were used. The globar source inside the FTS was used, together with a 6 µm thick mylar beam splitter, a band-pass filter, and a Si bolometer detector cooled down to 4.2 K by liquid helium. Measurements were made at room temperature (≈ 296 K, since the cell temperature cannot be varied) for a path length of 151.75 m, with an unapodized spectral resolution of 1 $cm^{-1}$. In total, 1000 scans were co-added to generate each of the spectra. Due to the optics and detector used, the signal is significant and reliable between 40 and 680 $cm^{-1}$. Within this range, each recording provides about 2650 spectral values with a step of 0.24 $cm^{-1}$. The pressures were measured and continuously monitored with both a 1000 millibar (Pfeifer) and a 1000 Torr (Edwards) gauges. Their readings agreed in all cases to better than 1 %. $CO_2$, $H_2$ and $CH_4$ gases provided by Air Liquide were used for the recording of pure samples as well as mixtures. The latter were made by first introducing $H_2$ (or $CH_4$) and then adding $CO_2$ up to a total pressure of about 0.95 bar. After waiting for about 1 hour to let the mixture homogenize a first spectrum was recorded, then the pressure of each initial gas sample was progressively reduced with a vacuum pump to yield spectra of the same mixture (or pure gas) at several lower pressures. In addition, reference (100% transmission) spectra were obtained for similar pressures by introducing pure argon into the cell. The list of recorded spectra and their conditions are given in Appendix A.

Note that we had to limit the total pressure in the cell to slightly below 1 bar in order to avoid possible damage of the cell and/or of its windows since they are not designed to go above atmospheric pressure. The product of the path length and gas densities is thus limited to about $1.3\ 10^4$ cm amagat$^2$ (1 amagat corresponds to $2.69\ 10^{19}$ molec/cm$^3$) for studies of pure gases and $0.33\ 10^4$ cm amagat$^2$ for mixtures, which induces large uncertainties on the determination of weak CIAs, as shown in Secs. 3 and 4. We also draw the reader's attention to the fact that the SOLEIL-AILES facility is extremely sought and busy and that we only had access to it for a few days, limiting the number of spectra we could record in this first experimental study of the CIAs by $H_2$-$CO_2$ and $CH_4$-$CO_2$ pairs. We hope that this will stimulate further thorough experimental investigations, hopefully under more absorbing conditions.

2.2   Spectra treatments

The raw spectra recordings were treated using the five steps procedure detailed below.

*Step 1:* For all samples, the recorded spectra show absorption by water vapor lines due to small amounts of $H_2O$ present in the gases used. The first step of the treatment of the collected data was thus to remove their contributions. For this, $H_2O$ transmissions were computed and used, as explained in Appendix B, in order to remove the absorption features due to water lines from the raw spectra and obtain "dried out" spectra.

*Step 2:* For a given sample, the transmission spectrum was then obtained by dividing the associated recording cleared from the $H_2O$ contribution by a pure argon spectrum, also cleared of the $H_2O$ lines. Among the various recorded pure Ar spectra (see Appendix A), we chose a linear combination of two Ar spectra for which the broad-band features resulting from the absorption and interference patterns generated by the propylene windows are the closest to those observed in the sample gas spectrum. Indeed, since these features are different from one spectrum to another (likely because of changes of the temperature and/or window curvature induced by pressure), ratioing adapted recordings is essential to minimize the residual broad-band oscillations in the resulting transmission spectrum. However, baseline variations due to changes of the globar source emission, bolometer responsivity and mechanical deformations of the multipass cell itself may also occur from one recording to another. This introduces an error on the 100% transmission level which directly affects the weak and broad-band absorption continua searched for. The way we dealt with this difficulty is described in step 5 below.

*Step 3:* For the part of this study devoted to the collision-induced absorption continua discussed in Sec. 3 (but not to the far line wings discussed in Sec. 4), the third step was to remove the contributions of the local lines of $CO_2$ and $CH_4$. These contributions were computed as explained in Appendix B, yielding the associated transmission by local lines by which the spectrum obtained from steps 1 and 2 is divided so that only the CIA remains.

*Step 4:* The transmissions obtained from the preceding treatments show a few sharp features (outliers) essentially found close to the local lines (of $H_2O$, $CH_4$ and $CO_2$) that have been removed in steps 1 and 3. Analysis shows that this is due to small differences between the measurements and the calculation of these narrow features. In order to "clean" the spectra from these outliers, we performed the following procedure. First, a moving average of each spectrum over 5 cm$^{-1}$ was computed. Then, the mean distance between this averaged spectrum and the original one was computed at each wavelength. All data points of the original spectrum that are further away from the averaged spectrum than half of this mean distance were rejected. Note that this 5 cm$^{-1}$-averaged spectrum is used only for the rejection of outliers and that the next steps of the analysis use the values of the original spectra that have not been rejected by this filtering.

*Step 5:* Starting from each transmission spectrum obtained from steps 1-4, we compute the negative value of its natural logarithm and divide the result by the optical path length (15175 cm). This provides the measured absorption coefficient $\alpha^{meas}(\sigma, P_X, P_{CO_2})$ at wave number $\sigma$ that, since local line contributions have been removed and only CIA remains, should be equal to:

$$\alpha^{calc}(\sigma, P_X, P_{CO_2}) = P_{CO_2}^2 C_{CO_2-CO_2}(\sigma) + P_X P_{CO_2} C_{CO_2-X}(\sigma) + P_X^2 C_{X-X}(\sigma) \quad , \quad (1)$$

where $C_{CO2-X}(\sigma)$ is the pressure-normalized binary collision-induced absorption coefficient for $CO_2$-X pairs. In order to determine its spectral values, two procedures have been used:

- In the first procedure, we assume that the 100% transmission baseline is correct. For pure gases, we determine $C_{X-X}(\sigma_i)$ from a simultaneous least square fit of all spectra, by minimizing the quantity $\sum_{P_X} \sum_{\sigma_i} \left[ \alpha^{meas}(\sigma_i, P_X) - P_X^2 C_{X-X}(\sigma_i) \right]^2$. For $CO_2$-X mixtures, we minimize

$$\sum_{P_X} \sum_{P_{CO_2}} \sum_{\sigma_i} \left\{ \left[ \alpha^{meas}(\sigma_i, P_X, P_{CO_2}) - P_{CO_2}^2 C_{CO_2-CO_2}(\sigma_i) - P_X^2 C_{X-X}(\sigma_i) \right] - \left[ P_X P_{CO_2} C_{CO_2-X}(\sigma_i) \right] \right\}^2$$

by floating the values of the $C_{CO_2-X}(\sigma_i)$, those of $C_{X-X}(\sigma_i)$ and $C_{CO_2-CO_2}(\sigma_i)$ being fixed. For the latter, we used the data of Ho et al. 1971 for pure $CO_2$, of Codastefano et al. 1985 for pure $CH_4$ and of Abel et al. 2011 for pure $H_2$, because they are more precise than the ones determined in the present study. These least square fits also provide the 1-$\sigma$ statistical (rms) uncertainties on the fitted parameters [$C_{X-X}(\sigma_i)$ and $C_{CO_2-X}(\sigma_i)$]. Note that 2-$\sigma$ was adopted for the error bars displayed in the figures below.

- In the second procedure, we introduce a baseline correction as a linear function of wave number. This function is assumed to be the same for a series of consecutive spectra at different pressures (i.e. those for a given mixture or a pure gas). In this case, for pure gases, we minimize the quantity $\sum_{P_X} \sum_{\sigma_i} \left[ \alpha^{meas}(\sigma_i, P_X) - A - B\sigma_i - P_X^2 C_{X-X}(\sigma_i) \right]^2$ by floating the values of A, B and $C_{X-X}(\sigma_i)$. For $CO_2$-X mixtures, since we recorded spectra at several pressures for each sample composition (i.e. each value of $P_X / P_{CO_2}$, see Table 1), we minimize the quantity

$$\sum_{P_X} \sum_{P_{CO_2}} \sum_{\sigma_i} \left\{ \left[ \alpha^{meas}(\sigma_i, P_X, P_{CO_2}) - P_{CO_2}{}^2 C_{CO_2-CO_2}(\sigma_i) - P_X{}^2 C_{X-X}(\sigma_i) \right] \right.$$

$$\left. - A(P_X/P_{CO_2}) - B(P_X/P_{CO_2})\sigma_i - \left[ P_X P_{CO_2} C_{CO_2-X}(\sigma_i) \right] \right\}^2$$

by floating the values of $A(P_X/P_{CO_2})$, $B(P_X/P_{CO_2})$, and $C_{CO_2-X}(\sigma_i)$. We thus assume here that A and B have a unique value for each mixture relative composition. Note that there are about 1400 points in each fitted spectrum and that, for each mixture, nine spectra (3 pressures for each one of the three mixtures, see Table 1) were recorded. We thus have a total number of ~12600 absorbance values. They provide enough information to deduce, from their adjustments, the ~1400 pressure-normalized absorptions $C_{CO_2-X}(\sigma_i)$ and the 6 baseline parameters $A(P_X/P_{CO2})$, $B(P_X/P_{CO2})$. However, since the retrieved values of $C_{CO_2-X}(\sigma_i)$ remain noisy and uncertain, a moving spectral average will be made in order to deduce the final values listed in Appendix C and plotted in Fig. 7, as explained at the end of Sec. 3.2.

**3. The Collision-Induced Absorptions (CIAs)**

3.1 The test cases of the CIAs of pure $CO_2$, $CH_4$ and $H_2$

Although measuring the CIAs of pure gases is not the objective of this study since several measurements and calculations have been made before (see below), we measured them to test and evaluate the accuracy of our experimental and data analysis procedures. Indeed, since the previous experimental studies have been made under much more favorable conditions (stronger absorptions thanks to higher pressures), they are likely more accurate than our determinations.

For the pure $CO_2$ CIA, measurements were reported in Harries 1970, Ho et al. 1971 and Hartmann et al. 2011, for instance, and theoretical calculations can be found in Gruszka & Borysow 1997, Gruszka & Borysow 1998 and Hartmann et al. 2011. A comparison of our results, obtained from fits of the five spectra at pressures from about 0.20 to 0.95 bar with and without introduction of a baseline, with the measured values of Ho et al. 1971 (uncertainty stated as better than 10 %) and the calculated results of Gruszka & Borysow 1997 is plotted in Fig. 1[(#)]. As can be seen, adjusting a baseline in the fit significantly reduces the error bars (and rms of the fit), corrects for the negative values in the high frequency wing, and improves the agreement between our results and the previous determinations. In this case, the area obtained from the integration of our results between 60 and 250 cm$^{-1}$ is $2.74 \pm 0.32 \ 10^{-3}$ cm$^{-2}$/amagat$^2$ (the uncertainty given here being directly obtained from the error bars in Fig. 1) while $3.11 \pm 0.31 \ 10^{-3}$ cm$^{-2}$/amagat$^2$ is obtained from the measurements of Ho et al. 1971, and $3.73 \ 10^{-3}$ cm$^{-2}$/amagat$^2$ from the calculated values of Gruszka & Borysow 1997. Our determination thus agrees, within uncertainties, with the previous measurement of Ho et al. 1971. Note that, when compared to the other gas samples discussed below, the pure $CO_2$ case is the most favorable since, for the most optically thick sample investigated (at ~0.95 bar), the transmission is 55% at the peak near 50 cm$^{-1}$.

---

[(#)] The Amagat density unit, generally used in studies of CIA, corresponds to the number density of molecules at $P_0$=1 atm ad $T_0$=273.15 K, i.e. Loschmidt's number $n_0$=2.686 $10^{19}$ molec/cm$^3$. For an ideal gas at pressure $P$(atm) and temperature $T$(K), the density in amagat is $(P/P_0)(T_0/T)$.

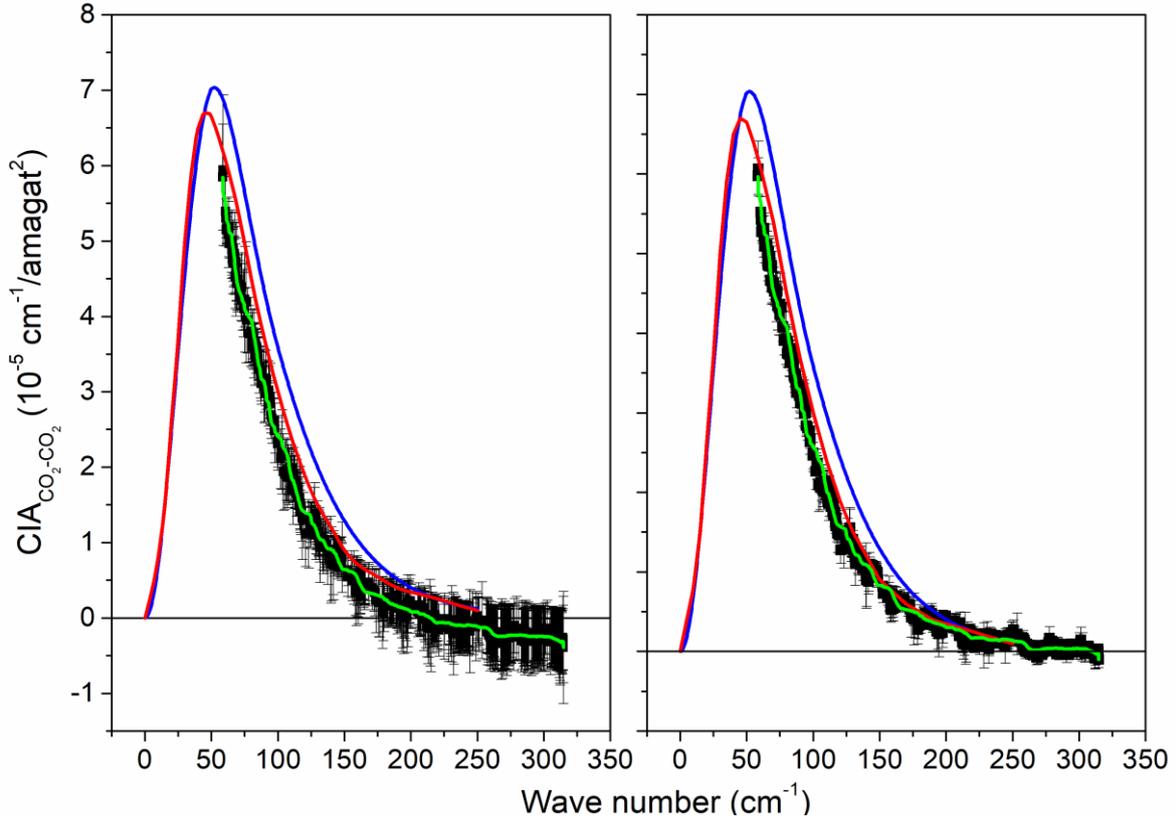

*Fig. 1: Pure $CO_2$ CIAs at room temperature. The symbols with error bars are the present measurements, the green line is a moving average of the latter over 10 $cm^{-1}$, the red line represents the measurements of Ho et al. 1971 (digitized from Fig. 1 of this reference), and the blue curve denotes the calculated values from Gruszka & Borysow 1997 which are available in the HITRAN CIA database section (Richard et al. 2012). The results in the left and right hand-side panels have been obtained without and with the baseline adjustment, respectively.*

The CIA of pure $CH_4$ was investigated experimentally in Codastefano et al. 1985, Codastefano et al. 1986, Birnbaum 1975 and Dagg et al. 1986. El-Kader & Maroulis 2012, Borysow & Frommhold 1987a and Borysow & Frommhold 1987b provide examples of theoretical calculations. A comparison between some of these previous data and the results obtained from our three spectra (pressures between ~0.60 and 0.98 bar) is plotted in Fig. 2. As for pure $CO_2$, introducing a baseline in the fit reduces the uncertainties (and rms), improves the $CH_4$-$CH_4$ CIA high frequency wing, and leads to a better agreement with previous determinations. In this case, the area obtained from the integration of our results between 70 and 600 $cm^{-1}$ is $2.2 \pm 0.2\ 10^{-3}$ $cm^{-2}/amagat^2$. It is slightly lower than the experimental value of $2.97 \pm 0.31\ 10^{-3}$ $cm^{-2}/amagat^2$ from Codastefano et al. 1985. Note that, for the most optically thick sample investigated (at ~0.98 bar), the peak absorption in our measurements is significantly smaller than for pure $CO_2$ with a transmission of about 90% near 250 $cm^{-1}$.

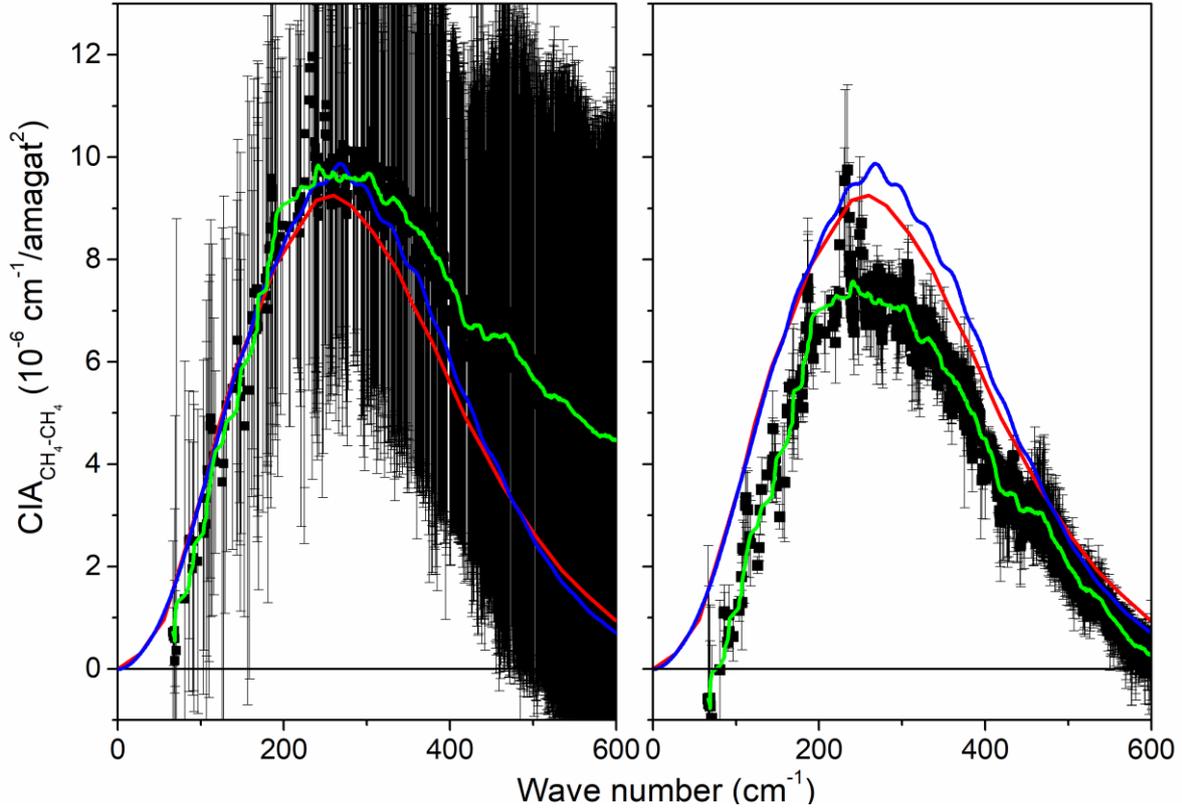

*Fig. 2: Pure CH$_4$ CIAs at room temperature. The symbols with error bars are the present measurements, the green line is a moving average of the latter over 10 cm$^{-1}$, the red line represents the measurements of Codastefano et al. 1985 (digitized from Fig. 1 of this reference), and the blue curve denotes the calculated values from Borysow & Frommhold 1987a which are available in the HITRAN CIA database section (Richard et al. 2012). The results in the left and right hand-side panels have been obtained without and with adjustment of the baseline, respectively.*

The CIA of pure H$_2$ was investigated experimentally in Birnbaum 1978 and Bachet et al. 1983. Theoretical calculations were made in Birnbaum et al. 1996, Gustafsson et al. 2003, Abel et al. 2011, Karman et al. 2015b, and references therein. A comparison between some of these previous data and the results of our measurements is plotted in Fig. 3. Note that in this case, the treatment applied to the spectra was slightly different from that used for all other samples. In fact, a quick look at the transmissions obtained after step 4 revealed that they differed from unity at small wave numbers (where the absorption should be negligible (Birnbaum 1978) for our recording conditions) with significant inconsistencies between the three spectra. This explains the very large uncertainties and negative values of the CIA obtained when no base line is introduced in the fit (see the left panel of Fig. 3). Therefore, we pre-treated each transmission spectrum by dividing it by the averaged value at about 100 cm$^{-1}$ where the absorbance under our conditions should be nearly zero. Then, the procedure described in step 5 of Sec. 2 was applied with the fit of a single base line for all spectra. This leads to the results plotted in the right hand side panel of Fig. 3 which show several obvious improvements with respect to those in the left hand side panel. However, significant differences with previous determinations remain. These differences can be likely explained by the fact that, for the most optically thick sample investigated (i.e. about 0.95 bar of H$_2$), the peak absorption is very small with a transmission of about 94% near 600 cm$^{-1}$ (according to the values of Birnbaum 1978).

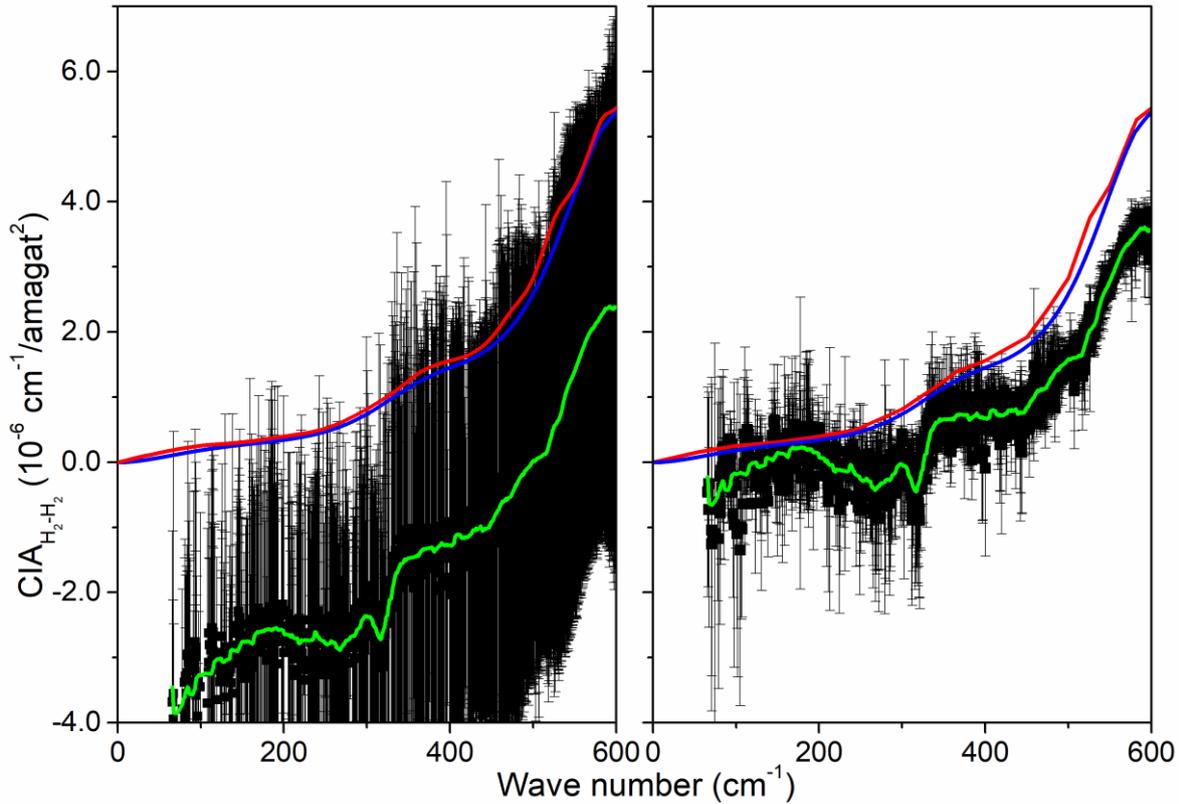

*Fig. 3: Pure $H_2$ CIA at room temperature. The symbols are the present measurements, the green line is a moving average of the latter over 10 cm$^{-1}$, the red line represents the measurements of Birnbaum 1978 (digitized from Fig. 2 of this reference), and the blue curve denotes the calculated values from Abel et al. 2011 which are available in the HITRAN CIA database section (Richard et al. 2012).*

3.2 The CIAs of $CH_4+CO_2$ and $H_2+CO_2$ mixtures

Now that the comparisons between our results for pure gases and previous determinations have validated our experimental and data analysis procedure (at least when the peak absorption is large enough, thus excluding the pure $H_2$ case) we consider the mixtures for which no other measurements are available. Note that, while squared pressures up to ~1 bar$^2$ could be used for pure gases, the case of mixtures is much less favorable since $P_X P_{CO_2}$ is now limited to ~0.25 bar$^2$. Finally, since the results of Sec. 2 show that adjusting a linear baseline correction leads to significant improvements, only the results obtained with this approach are presented below. However, for the two considered mixtures, results obtained without adjustment of baselines are fully consistent when error bars are considered.

The CIA of $CH_4+CO_2$ has never been measured before, but theoretical predictions have been made (Wordsworth et al. 2017). A comparison between these predictions and the results of our measurements is plotted in Fig. 4. We measure a significant $CO_2$-$CH_4$ CIA in the 50-500 cm$^{-1}$ spectral range that has, as predicted (Wordsworth et al. 2017), two spectral regions of higher absorption around 50 and 270 cm$^{-1}$. However, our experiments lead, on average, to significantly smaller values with an area between 50 and 500 cm$^{-1}$ of 9.9±2.7 10$^{-3}$ cm$^{-2}$/amagat$^2$, about 1.7 times smaller than that obtained from the prediction of Wordsworth et al. 2017. Despite the rather large uncertainties, the experimental results show that the theoretical prediction of Wordsworth et al. 2017 overestimates this CIA. Indeed, our CH$_4$+CO$_2$ CIA leads to peak absorbances for the optically thickest conditions investigated (~0.47 bar of CH$_4$ + ~0.47 bar of CO$_2$) of 0.16 and 0.10, corresponding to a transmission of about 85 % and 90 % near 50 and 270 cm$^{-1}$ respectively, which gives relatively "comfortable" conditions.

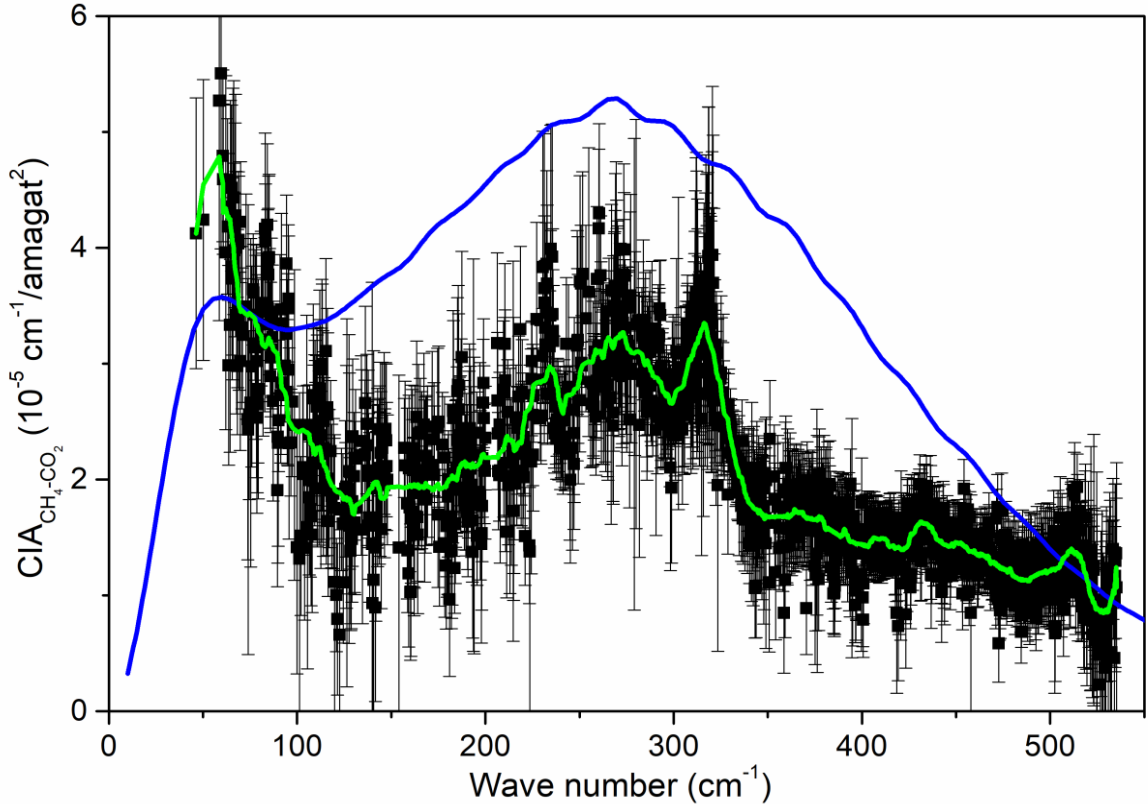

*Fig. 4:* CH$_4$+CO$_2$ *CIAs at room temperature. The symbols with error bars are the present measurements, the green line is a moving average of the latter over 10 cm$^{-1}$, the blue line represents the calculated values of Wordsworth et al. 2017.*

As for the CIA of CH$_4$+CO$_2$, that of H$_2$+CO$_2$ has never been measured before, but theoretical predictions have also been made (Wordsworth et al. 2017). These calculated data and the results of our experiments are plotted in Fig. 5. We measure a significant CIA throughout the entire spectral range (corresponding to an absorbance of ~0.07 for the optically thickest conditions) which increases with wavenumber at high frequencies, as predicted in Wordsworth et al. 2017. However, large differences can be observed in Fig. 5 between our measurements and the predictions, both for the shape and magnitude, for which we have no explanation at this step. Finally note that our determinations are limited to wavenumbers lower than 550 cm$^{-1}$ because of the strong contribution of CO$_2$ absorption lines beyond this frequency (see Fig. 6). In addition, it is possible that a part of the absorption that we measure comes from the H$_2$-broadened wings of the 15 μm CO$_2$ band lines, whose contributions are, as the CIA, proportional to $P_{H_2}P_{CO_2}$.

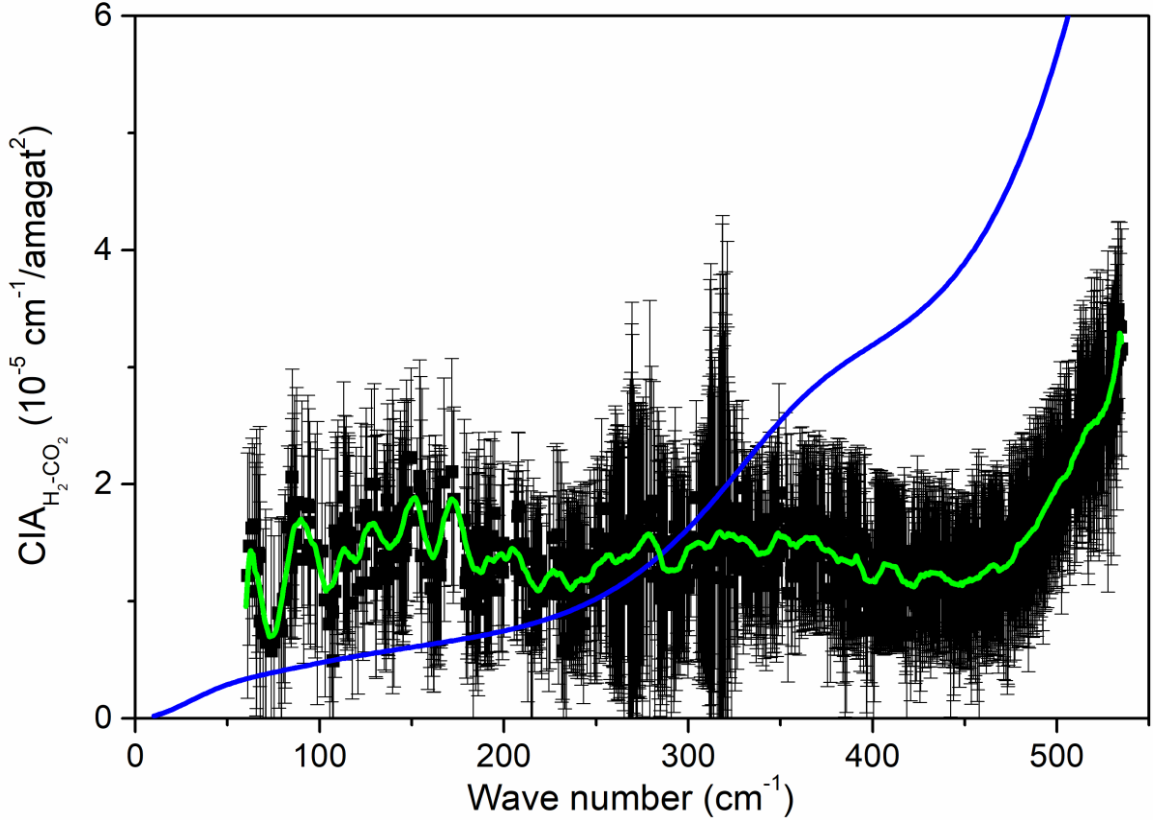

*Fig. 5:* H$_2$+CO$_2$ *CIAs at room temperature. The symbols with error bars are the present measurements, the green line is a moving average of the latter over 10 cm$^{-1}$, the blue line represents the calculated values of Wordsworth et al. 2017.*

In order to generate "the best" datasets of measured CH$_4$+CO$_2$ and H$_2$+CO$_2$ CIAs we started from the values with error bars in Figs. 4 and 5. For the mean values, we computed a moving average over 25 cm$^{-1}$ and tabulated its values with a step of 5 cm$^{-1}$. Determining the uncertainties is, as it is well known, a very difficult task, even in cases where numerous measurements have been made for reliable statistics. In the present study, we computed the moving average, over 25 cm$^{-1}$, of the error bars plotted in Figs. 4 and 5. The obtained results were then multiplied by 1.5 in order to take into account other sources of uncertainties (mostly due to base line stability within the same recording series). Since the error bars in Figs. 4 and 5 already correspond to the 2-σ value (95% confidence interval), we believe that the total uncertainty estimated this way is (very) conservative considering the measured data at our disposal. The final set of data for the CH$_4$+CO$_2$ and H$_2$+CO$_2$ CIAs are given in Table 2 of Appendix C. We recall the reader that these should be used with great care as their uncertainties are rather large.

## 4. The low frequency wing of the CO$_2$ ν$_2$ band

The recorded pure CO$_2$ spectra enable to study the absorption in the low frequency wing of the intense ν$_2$ band (centered at 667 cm$^{-1}$), from about 500 to 600 cm$^{-1}$. While there are, to our knowledge, no previous measurements available, comparisons with calculations are possible. Indeed, the spectra can be predicted using the line-mixing model of Tran et al. 2011 or Voigt profiles corrected in the wings by using a χ factor. Three sets of such factors have been proposed, deduced from measurements in the high frequency wings of the ν$_2$ (σ>750

cm$^{-1}$) (Tran et al. 2011) and $\nu_3$ ($\sigma$>2400 cm$^{-1}$) (Perrin & Hartmann 1989,Tran et al. 2011) bands. Note that the line-mixing model of Tran et al. 2011 and the $\chi$ factor of Perrin & Hartmann 1989 were used in Turbet & Tran 2017 for radiative transfer calculations in the far infrared for rich $CO_2$ atmospheres but that they had not been validated in this spectral region yet. Comparisons between measured transmissions and those computed with the four models mentioned above and by using purely Voigt profiles (no line-mixing and no $\chi$-factor correction, i.e. $\chi(\Delta\sigma) = 1$) are plotted in Fig. 6, calling for the following remarks. The first is that calculations with pure Voigt line shapes (blue) lead to very large errors showing the strongly sub-Lorentzian behavior of the line wings. The second is that the three $\chi$-factors used (red curves) lead to indiscernible results on this plot, all in very good agreement with the measured spectrum, except in a narrow interval around the Q branch centered at 545 cm$^{-1}$. This local effect of line-mixing is very nicely reproduced by the line-mixing model of Tran et al. 2011 which is also in excellent agreement with the $\chi$-factor predictions (as also shown in Turbet & Tran 2017) and the measured spectra throughout the entire spectral range.

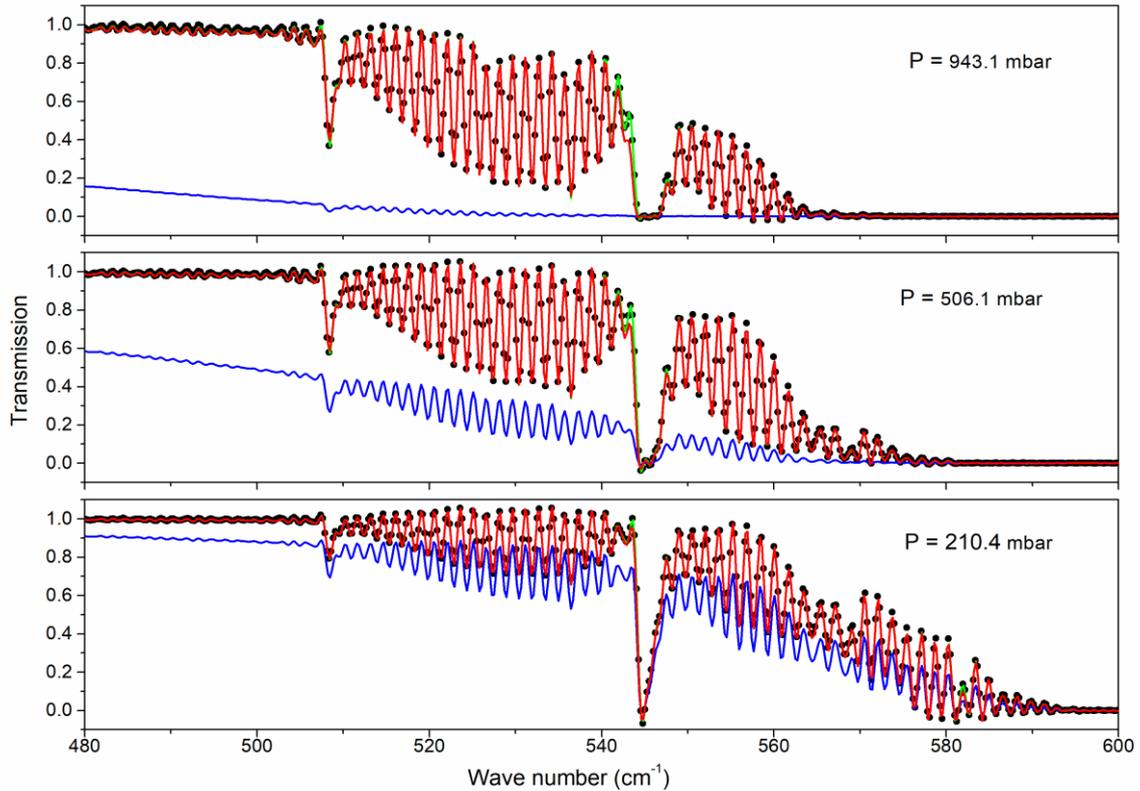

*Fig. 6: Pure $CO_2$ transmission spectra for three pressures. The black circle denotes the measured values (not all plotted). The red lines are the (indistinguishable) results obtained with the three $\chi$-factors (see text). The blue line is obtained with purely Voigt line shapes ($\chi = 1$). The green line denotes the results of calculations obtained with the pure $CO_2$ line-mixing model of Tran et al. 2011.*

## 5. Discussion

5.1 On the CIAs of $CH_4+CO_2$ and $H_2+CO_2$

A comparison between our final experimental determinations of the $CH_4+CO_2$ and $H_2+CO_2$ CIAs and the corresponding predictions of Wordsworth et al. 2017 as well as with the CIAs when $CO_2$ is replaced by $N_2$ is plotted in Fig. 7.

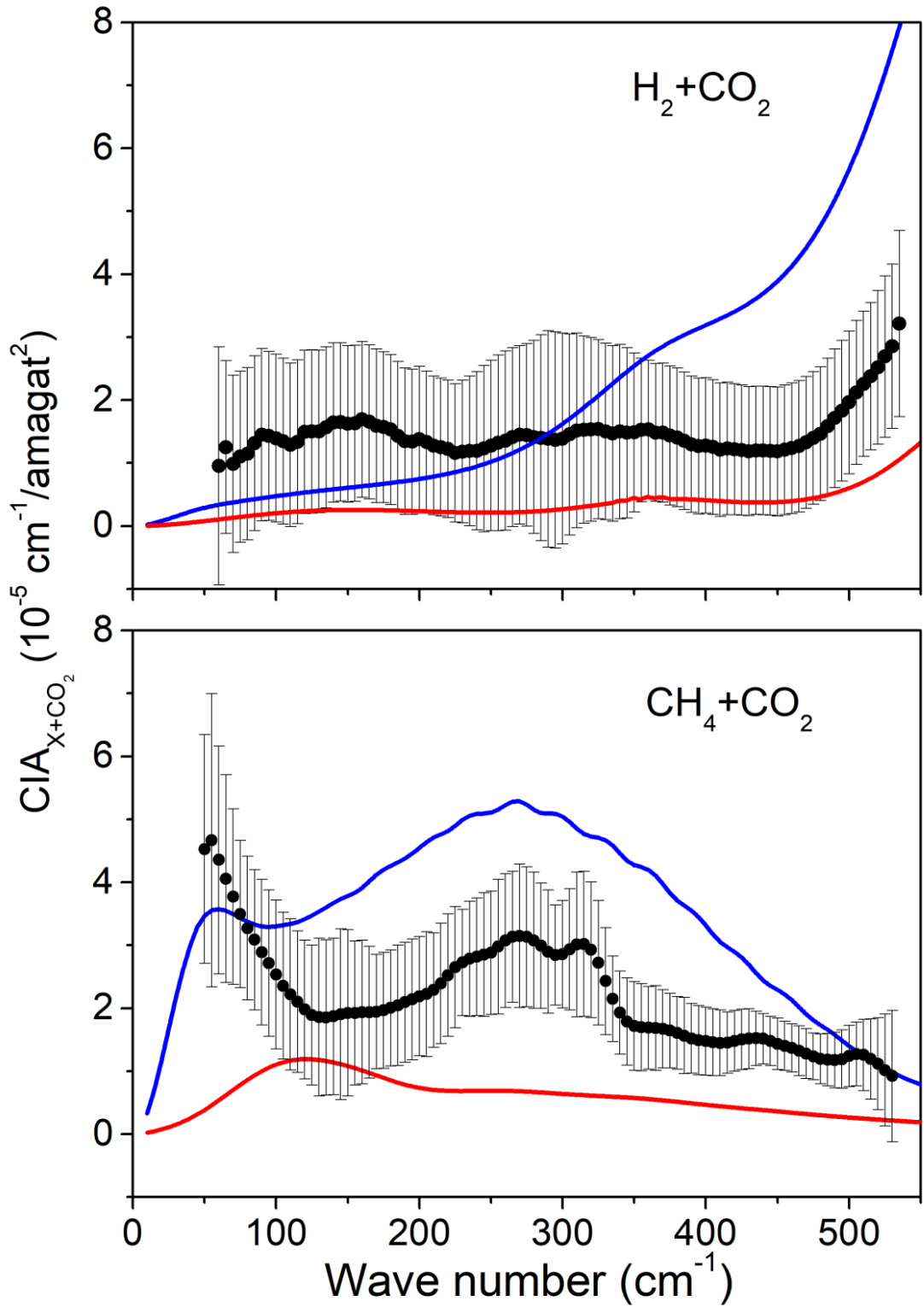

*Fig. 7: In black, $H_2+CO_2$ CIA (top panel) and $CH_4+CO_2$ CIA (bottom panel) measured at room temperature (symbols with error bars). The blue lines are the calculated valued of $CO_2$-$H_2$ and $CO_2$-$CH_4$, respectively, taken from Wordsworth et al. 2017. The red lines are the CIA of $N_2$-$H_2$ (Borysow et al. 1986) and $N_2$-$CH_4$ (Borysow & Tang 1993), respectively, taken from the HITRAN CIA database section (Richard et al. 2012).*

First, both our measurements and the predictions indicate that the effect of interactions of $CH_4$ and $H_2$ with $CO_2$ induce much stronger absorptions than with $N_2$. This could actually be expected by simply considering the long range induction mechanisms which are likely dominant here and are governed by the isotropic polarizability and the electric quadrupole moment (of $N_2$ or $CO_2$). As shown by the data given in Li et al. 1998, the values of all these parameters are, for $CO_2$, from 1.5 to nearly 3 times greater than for $N_2$. Reminding that the square of the dipole is involved, and even though the intermolecular potential and its anisotropy also play a role, it is quite obvious that the X-$CO_2$ should be significantly larger than the X-$N_2$ CIA.

For the differences between our measurements and the predictions of Wordsworth et al. 2017, which are significant and concern both the magnitude and the shape of the CIAs, we have no explanation at this step. Trying to understand their origin(s) is beyond the scope of the present paper.

Obviously, independent checks of our results by new measurements are needed, if possible under more absorbing conditions in order to reduce the uncertainties resulting from a poor knowledge of the 100% transmission level. Ideally, we recommend performing future investigations of $CO_2$-$CH_4$ and $CO_2$-$H_2$ CIAs for products of the path length and gas densities up to at least $2\ 10^4$ cm amagat$^2$. Furthermore, measurements at several temperatures (between at least 200 and 300 K) would be of considerable interest since they would provide the planetary science community with CIAs that can be used directly in numerical climate models. However, such experiments are difficult and there are not many (if any) existing set-ups suitable to carry them.

Note that the predictions of Wordsworth et al. 2017 were obtained using a very simple approach based on a combination of pure gas CIA data and not on direct calculations of the absorption spectrum of the mixture (detailed in their supporting information, Section 1.1.). Therefore, their quality needs to be investigated and this could be done by using various alternative theoretical approaches. These include fully classical computations based on molecular dynamics simulations (e.g. Gruszka & Borysow 1998, Hartmann et al. 2011, Hartmann et al. 2018a), semi-classical approaches based on line shapes (e.g. Birnbaum & Cohen 1976, Hunt & Poll 1978) and the so-called "isotropic approximation" (e.g. Leforestier et al. 2010, Hartmann et al. 2018a), and fully quantum scattering computations assuming isotropic intermolecular forces (e.g. Borysow et al. 1986, Borysow & Tang 1993) or taking the intermolecular potential anisotropy into account (e.g. Karman et al. 2015a, Karman et al. 2015b). Note that, in the absence of relevant measurements, theory may give information on the temperature dependences of the $CO_2$-$CH_4$ and $CO_2$-$H_2$ CIAs which are, as mentioned above, crucially needed.

5.2. Implications for the climate of early Mars

Although our measurements have large uncertainties, we can discuss their implications for the climate of early Mars. In the 200-600 cm$^{-1}$ spectral window of early Mars, our measurements confirm that the $CO_2$-$CH_4$ and $CO_2$-$H_2$ CIA are much stronger than those of $N_2$-$CH_4$ and $N_2$-$H_2$ pairs, respectively. However, they also show that the calculations of Wordsworth et al. 2017 overestimate on average the $CO_2$-$CH_4$ and $CO_2$-$H_2$ CIAs by a factor of about 1.7 and 1.6, respectively.

Our measurements indicate that Wordsworth et al. 2017 and Ramirez 2017 overestimated the effect of these CIAs in their numerical climate simulations and that they therefore underestimated the amount of $H_2$ and $CH_4$ required to warm the surface of early

Mars above the melting point of water, by a factor up to 2-3. Similarly, calculations of the boundaries of the Habitable Zone based on hydrogen (Ramirez & Kaltenegger 2017) or methane (Ramirez & Kaltenegger 2018) should be revised accordingly. Future experiments and calculations of these CIAs combined with numerical climate model simulations will eventually and hopefully better constrain these numbers.

## 6. Conclusion

We performed the first measurements of the far infrared CIAs of $CO_2$-$CH_4$ and $CO_2$-$H_2$ mixtures as well as of the absorption by the (low frequency) wings of the pure $CO_2$ 15 µm band lines. We confirm the theoretical prediction of Wordsworth et al. 2017 that the $H_2$+$CO_2$ and $CH_4$+$CO_2$ CIAs are significantly stronger in the 40-600 cm$^{-1}$ region than those of $H_2$+$N_2$ and $CH_4$+$N_2$, respectively. However, our results for $H_2$+$CO_2$ and $CH_4$+$CO_2$ significantly differ from the predictions, both in terms of the magnitude and shape of the CIAs. In the expected 200-600 cm$^{-1}$ spectral window of early Mars, our measurements show that the calculations of Wordsworth et al. 2017 overestimate in average the $CO_2$-$CH_4$ and $CO_2$-$H_2$ CIAs by a factor of about 1.7 and 1.6, respectively. The amount of $H_2$ and $CH_4$ required to warm the surface of early Mars above the melting point of water should be revised accordingly. For the pure $CO_2$ 15 µm-band line-wings, we show that both the χ-factor deduced from measurements near 4 µm and a line-mixing model very well describe the observed strongly sub-Lorentzian behavior in the 500-600 cm$^{-1}$ region. The results of this first experimental investigation obviously need confirmation by independent measurements and we hope that the results presented in this paper will stimulate new laboratory investigations. Similarly, it would be of great interest to compare the results of our measurements and of the only prediction available with those obtained from alternative models and calculations.

**Acknowledgements**
The authors thank R. Wordsworth for providing the calculated $CH_4$+$CO_2$ and $H_2$+$CO_2$ CIA spectra plotted in Fig. 2 of the Supporting Information file of Ref. (Wordsworth et al. 2017). They are grateful to R. Gamache for providing the data for the broadening of $H_2O$ lines by $CO_2$. They also thank the SOLEIL-AILES beamline staff for their technical support during this project and C. Janssen and L. Manceron for lending pressure transducers.

# Appendix A – Conditions of the recorded spectra

We performed various measurements of transmission spectra of Ar, $CO_2$, $CH_4$, $H_2$ as well as of $H_2+CO_2$ and $CH_4+CO_2$ mixtures at room temperature (23 °C). In total, we recorded 39 spectra:
- 7 spectra of pure Argon between 100 and 950 mbar.
- 5 spectra of pure $CO_2$ between 200 and 950 mbar.
- 3 spectra of pure $CH_4$ between 600 and 980 mbar.
- 3 spectra of pure $H_2$ between 600 and 950 mbar.
- 9 spectra of $CO_2+CH_4$ between 500 and 950 mbar, for three different mixtures (30,50 and 70% of $CH_4$).
- 10 spectra of $CO_2+H_2$ between 400 and 950 mbar, for three different mixtures (30,50 and 70% of $H_2$).

The experimental conditions of the recorded spectra are detailed in Table 1.

| Spectrum | $P_{Tot}$ (mbar) | $x_{Ar}$ (%) | $X_{H2}$ (%) | $X_{CH4}$ (%) | $X_{CO2}$ (%) |
|---|---|---|---|---|---|
| S_Ar-Ar#1 | 94.9 | 100 | 0.0 | 0.0 | 0.0 |
| S_Ar-Ar#2 | 207.2 | 100 | 0.0 | 0.0 | 0.0 |
| S_Ar-Ar#3 | 380.9 | 100 | 0.0 | 0.0 | 0.0 |
| S_Ar-Ar#4 | 496.1 | 100 | 0.0 | 0.0 | 0.0 |
| S_Ar-Ar#5 | 649.7 | 100 | 0.0 | 0.0 | 0.0 |
| S_Ar-Ar#6 | 751.2 | 100 | 0.0 | 0.0 | 0.0 |
| S_Ar-Ar#7 | 935.6 | 100 | 0.0 | 0.0 | 0.0 |
| S_CO2-CO2#1 | 210.5 | 0.0 | 0.0 | 0.0 | 100 |
| S_CO2-CO2#2 | 329.0 | 0.0 | 0.0 | 0.0 | 100 |
| S_CO2-CO2#3 | 506.4 | 0.0 | 0.0 | 0.0 | 100 |
| S_CO2-CO2#4 | 707.7 | 0.0 | 0.0 | 0.0 | 100 |
| S_CO2-CO2#5 | 942.8 | 0.0 | 0.0 | 0.0 | 100 |
| S_CH4-CH4#1 | 610.8 | 0.0 | 0.0 | 100 | 0.0 |
| S_CH4-CH4#2 | 751.3 | 0.0 | 0.0 | 100 | 0.0 |
| S_CH4-CH4#3 | 978.5 | 0.0 | 0.0 | 100 | 0.0 |
| S_H2-H2#1 | 597.1 | 0.0 | 100 | 0.0 | 0.0 |
| S_H2-H2#2 | 749.6 | 0.0 | 100 | 0.0 | 0.0 |
| S_H2-H2#3 | 947.2 | 0.0 | 100 | 0.0 | 0.0 |
| S_CO2-CH4#1 | 495.2 | 0.0 | 0.0 | 32.33 | 67.67 |
| S_CO2-CH4#2 | 634.3 | 0.0 | 0.0 | 32.33 | 67.67 |
| S_CO2-CH4#3 | 929.5 | 0.0 | 0.0 | 32.33 | 67.67 |
| S_CO2-CH4#4 | 601.5 | 0.0 | 0.0 | 49.88 | 50.12 |
| S_CO2-CH4#5 | 748.2 | 0.0 | 0.0 | 49.88 | 50.12 |
| S_CO2-CH4#6 | 939.8 | 0.0 | 0.0 | 49.88 | 50.12 |
| S_CO2-CH4#7 | 602.3 | 0.0 | 0.0 | 73.70 | 26.30 |
| S_CO2-CH4#8 | 754.0 | 0.0 | 0.0 | 73.70 | 26.30 |
| S_CO2-CH4#9 | 929.5 | 0.0 | 0.0 | 73.70 | 26.30 |
| S_CO2-H2#1 | 400.6 | 0.0 | 49.42 | 0.0 | 50.58 |
| S_CO2-H2#2 | 601.8 | 0.0 | 49.42 | 0.0 | 50.58 |
| S_CO2-H2#3 | 754.2 | 0.0 | 49.42 | 0.0 | 50.58 |
| S_CO2-H2#4 | 945.1 | 0.0 | 49.42 | 0.0 | 50.58 |
| S_CO2-H2#5 | 537.0 | 0.0 | 31.91 | 0.0 | 68.09 |
| S_CO2-H2#6 | 758.3 | 0.0 | 31.91 | 0.0 | 68.09 |
| S_CO2-H2#7 | 947.0 | 0.0 | 31.91 | 0.0 | 68.09 |

| S_CO2-H2#8  | 600.4 | 0.0 | 68.89 | 0.0 | 31.11 |
| S_CO2-H2#9  | 747.7 | 0.0 | 68.89 | 0.0 | 31.11 |
| S_CO2-H2#10 | 927.0 | 0.0 | 68.89 | 0.0 | 31.11 |

Table 1: Conditions of the recorded spectra.

# Appendix B – removing the local lines of $H_2O$, $CO_2$ and $CH_4$

In order to remove the contribution of $H_2O$ lines from the raw recorded spectra, the absorption coefficient of this species under the temperature, pressure and mixture compositions of the measurements was computed using Voigt line shapes with a cut-off 3 cm$^{-1}$ away from the line center, using the kspectrum code (Eymet et al. 2016). We used the line positions and intensities from the 2012 edition of the HITRAN database (Rothman et al. 2013). For the pressure-broadening coefficients, we used data from Brown et al. 2007 for $H_2O$-$CO_2$, a factor of 1.14 applied to the air-broadening coefficients from HITRAN for $H_2O$-$CH_4$ (Nwaboh et al. 2014). For $H_2$-broadening coefficients of $H_2O$ lines, a factor of 1 was used (Steyert et al. 2004), as well as for Ar- broadening. Note that the accuracy of the broadening coefficients used is not crucial since the spectral resolution of the measured spectra (1 cm$^{-1}$) is much larger than the line half-width (about 0.1 cm$^{-1}$ at 1 atm). The transmission was then computed and convolved by the Fourier transform spectrometer instrument line shape (a cardinal sine function with a resolution of 1 cm$^{-1}$) and the relative $H_2O$ amount was floated until the best agreement between measured and calculated absorptions around $H_2O$ lines was obtained. The measured spectrum was then corrected by dividing it by the computed transmission of water vapor. Note that the retrieved $H_2O$ relative amounts are small (between 1.2 10$^{-6}$ and 8.0 10$^{-6}$ mol/mol), making the contributions of the self and foreign water vapor continua (Mlawer et al. 2012, Ma & Tipping 1992) fully negligible here.

The same approach was used in order to remove the absorption by local $CH_4$ and $CO_2$ lines from the transmissions deduced from the spectra after their "cleaning" from the water vapor lines. $CO_2$ and $CH_4$ local lines were computed with the kspectrum code (Eymet et al. 2016) by using the HITRAN 2012 database (Rothman et al. 2013) and Voigt line shapes, with a cut-off 3 cm$^{-1}$ away from the line center. We used the self and air-broadening coefficients from HITRAN 2012 for all mixtures, except for the pressure broadening of $CH_4$ by $CO_2$, where we applied a factor of 1.3 on the air-broadened Lorentzian half width at half maximum (deduced from Fissiaux et al. 2014 and Rothman et al. 2013). We used the known partial pressures of $CH_4$ and $CO_2$ to remove the local contributions of $CH_4$ and $CO_2$ lines from the recorded spectra because this yields a very good result. This good match validates the procedure used to make the gas mixtures and determine their compositions.

We did not remove $H_2$ quadrupole transitions from our recorded transmission spectra because they make a totally negligible contribution to our measured spectra because of their very small intensities and of the 1 cm$^{-1}$ spectral resolution used.

# Appendix C – Measured $CH_4$+$CO_2$ and $H_2$+$CO_2$ CIAs

| $H_2$-$CO_2$ CIA | | | $CH_4$-$CO_2$ CIA | | |
|---|---|---|---|---|---|
| Wavenumber (cm$^{-1}$) | Abs. coef. (cm$^{-1}$/amagat$^{-2}$) | Uncertainty (cm$^{-1}$/amagat$^{-2}$) | Wavenumber (cm$^{-1}$) | Abs. coef. (cm$^{-1}$/amagat$^{-2}$) | Uncertainty (cm$^{-1}$/amagat$^{-2}$) |
|     |          |          | 50  | 4.53E-05 | 1.82E-05 |
|     |          |          | 55  | 4.67E-05 | 2.33E-05 |
| 60  | 9.55E-06 | 1.89E-05 | 60  | 4.36E-05 | 1.81E-05 |
| 65  | 1.25E-05 | 1.37E-05 | 65  | 4.06E-05 | 1.65E-05 |
| 70  | 9.85E-06 | 1.41E-05 | 70  | 3.77E-05 | 1.39E-05 |
| 75  | 1.10E-05 | 1.36E-05 | 75  | 3.50E-05 | 1.17E-05 |
| 80  | 1.15E-05 | 1.37E-05 | 80  | 3.27E-05 | 1.14E-05 |
| 85  | 1.31E-05 | 1.39E-05 | 85  | 3.09E-05 | 1.12E-05 |
| 90  | 1.45E-05 | 1.37E-05 | 90  | 2.89E-05 | 1.16E-05 |
| 95  | 1.43E-05 | 1.34E-05 | 95  | 2.72E-05 | 1.16E-05 |
| 100 | 1.39E-05 | 1.34E-05 | 100 | 2.54E-05 | 1.18E-05 |
| 105 | 1.34E-05 | 1.32E-05 | 105 | 2.36E-05 | 1.17E-05 |
| 110 | 1.29E-05 | 1.30E-05 | 110 | 2.22E-05 | 1.20E-05 |
| 115 | 1.34E-05 | 1.30E-05 | 115 | 2.10E-05 | 1.12E-05 |
| 120 | 1.49E-05 | 1.30E-05 | 120 | 1.98E-05 | 1.10E-05 |
| 125 | 1.50E-05 | 1.30E-05 | 125 | 1.89E-05 | 1.11E-05 |
| 130 | 1.50E-05 | 1.29E-05 | 130 | 1.86E-05 | 1.25E-05 |
| 135 | 1.57E-05 | 1.28E-05 | 135 | 1.85E-05 | 1.23E-05 |
| 140 | 1.64E-05 | 1.26E-05 | 140 | 1.87E-05 | 1.25E-05 |
| 145 | 1.65E-05 | 1.25E-05 | 145 | 1.90E-05 | 1.36E-05 |
| 150 | 1.62E-05 | 1.25E-05 | 150 | 1.92E-05 | 1.32E-05 |
| 155 | 1.63E-05 | 1.24E-05 | 155 | 1.93E-05 | 1.14E-05 |
| 160 | 1.69E-05 | 1.24E-05 | 160 | 1.94E-05 | 1.14E-05 |
| 165 | 1.65E-05 | 1.22E-05 | 165 | 1.94E-05 | 1.05E-05 |
| 170 | 1.59E-05 | 1.22E-05 | 170 | 1.95E-05 | 9.11E-06 |
| 175 | 1.57E-05 | 1.20E-05 | 175 | 1.97E-05 | 9.21E-06 |
| 180 | 1.53E-05 | 1.19E-05 | 180 | 2.01E-05 | 9.22E-06 |
| 185 | 1.43E-05 | 1.18E-05 | 185 | 2.04E-05 | 9.58E-06 |
| 190 | 1.35E-05 | 1.16E-05 | 190 | 2.10E-05 | 9.52E-06 |
| 195 | 1.34E-05 | 1.15E-05 | 195 | 2.14E-05 | 9.59E-06 |
| 200 | 1.38E-05 | 1.15E-05 | 200 | 2.19E-05 | 9.48E-06 |
| 205 | 1.34E-05 | 1.12E-05 | 205 | 2.22E-05 | 9.91E-06 |
| 210 | 1.28E-05 | 1.10E-05 | 210 | 2.29E-05 | 9.09E-06 |
| 215 | 1.25E-05 | 1.10E-05 | 215 | 2.39E-05 | 9.56E-06 |
| 220 | 1.22E-05 | 1.09E-05 | 220 | 2.52E-05 | 9.71E-06 |
| 225 | 1.16E-05 | 1.10E-05 | 225 | 2.65E-05 | 9.63E-06 |
| 230 | 1.18E-05 | 1.13E-05 | 230 | 2.73E-05 | 8.43E-06 |
| 235 | 1.19E-05 | 1.19E-05 | 235 | 2.78E-05 | 9.14E-06 |
| 240 | 1.19E-05 | 1.25E-05 | 240 | 2.82E-05 | 9.32E-06 |
| 245 | 1.23E-05 | 1.32E-05 | 245 | 2.85E-05 | 9.82E-06 |
| 250 | 1.28E-05 | 1.35E-05 | 250 | 2.88E-05 | 9.78E-06 |
| 255 | 1.32E-05 | 1.39E-05 | 255 | 2.98E-05 | 1.07E-05 |
| 260 | 1.36E-05 | 1.42E-05 | 260 | 3.07E-05 | 1.06E-05 |
| 265 | 1.41E-05 | 1.42E-05 | 265 | 3.13E-05 | 1.04E-05 |
| 270 | 1.45E-05 | 1.42E-05 | 270 | 3.15E-05 | 1.14E-05 |
| 275 | 1.45E-05 | 1.45E-05 | 275 | 3.13E-05 | 1.11E-05 |
| 280 | 1.41E-05 | 1.53E-05 | 280 | 3.08E-05 | 1.08E-05 |
| 285 | 1.40E-05 | 1.64E-05 | 285 | 2.99E-05 | 9.96E-06 |
| 290 | 1.38E-05 | 1.72E-05 | 290 | 2.90E-05 | 9.79E-06 |
| 295 | 1.37E-05 | 1.72E-05 | 295 | 2.85E-05 | 7.97E-06 |
| 300 | 1.39E-05 | 1.67E-05 | 300 | 2.86E-05 | 8.53E-06 |
| 305 | 1.45E-05 | 1.60E-05 | 305 | 2.94E-05 | 9.63E-06 |
| 310 | 1.52E-05 | 1.55E-05 | 310 | 3.01E-05 | 1.15E-05 |
| 315 | 1.53E-05 | 1.49E-05 | 315 | 3.02E-05 | 1.15E-05 |
| 320 | 1.53E-05 | 1.46E-05 | 320 | 2.93E-05 | 1.08E-05 |
| 325 | 1.54E-05 | 1.42E-05 | 325 | 2.72E-05 | 9.91E-06 |
| 330 | 1.50E-05 | 1.40E-05 | 330 | 2.43E-05 | 8.50E-06 |
| 335 | 1.47E-05 | 1.40E-05 | 335 | 2.15E-05 | 6.80E-06 |
| 340 | 1.50E-05 | 1.38E-05 | 340 | 1.93E-05 | 6.59E-06 |
| 345 | 1.48E-05 | 1.35E-05 | 345 | 1.79E-05 | 6.94E-06 |
| 350 | 1.49E-05 | 1.26E-05 | 350 | 1.72E-05 | 7.07E-06 |
| 355 | 1.53E-05 | 1.14E-05 | 355 | 1.70E-05 | 6.93E-06 |
| 360 | 1.53E-05 | 1.09E-05 | 360 | 1.69E-05 | 6.58E-06 |
| 365 | 1.48E-05 | 1.09E-05 | 365 | 1.69E-05 | 6.71E-06 |
| 370 | 1.48E-05 | 1.10E-05 | 370 | 1.67E-05 | 6.23E-06 |
| 375 | 1.44E-05 | 1.10E-05 | 375 | 1.65E-05 | 5.93E-06 |
| 380 | 1.41E-05 | 1.10E-05 | 380 | 1.61E-05 | 5.74E-06 |
| 385 | 1.35E-05 | 1.09E-05 | 385 | 1.56E-05 | 6.05E-06 |
| 390 | 1.30E-05 | 1.08E-05 | 390 | 1.52E-05 | 5.52E-06 |
| 395 | 1.27E-05 | 1.08E-05 | 395 | 1.49E-05 | 5.50E-06 |
| 400 | 1.28E-05 | 1.07E-05 | 400 | 1.48E-05 | 5.41E-06 |
| 405 | 1.25E-05 | 1.07E-05 | 405 | 1.46E-05 | 5.31E-06 |
| 410 | 1.21E-05 | 1.06E-05 | 410 | 1.45E-05 | 5.21E-06 |
| 415 | 1.23E-05 | 1.05E-05 | 415 | 1.45E-05 | 4.88E-06 |
| 420 | 1.22E-05 | 1.04E-05 | 420 | 1.48E-05 | 4.82E-06 |
| 425 | 1.20E-05 | 1.03E-05 | 425 | 1.50E-05 | 4.80E-06 |
| 430 | 1.19E-05 | 1.02E-05 | 430 | 1.52E-05 | 4.78E-06 |
| 435 | 1.20E-05 | 1.02E-05 | 435 | 1.52E-05 | 4.31E-06 |
| 440 | 1.20E-05 | 1.02E-05 | 440 | 1.52E-05 | 4.00E-06 |
| 445 | 1.19E-05 | 1.02E-05 | 445 | 1.48E-05 | 3.92E-06 |
| 450 | 1.18E-05 | 1.03E-05 | 450 | 1.43E-05 | 4.10E-06 |
| 455 | 1.21E-05 | 1.03E-05 | 455 | 1.40E-05 | 3.92E-06 |
| 460 | 1.23E-05 | 1.04E-05 | 460 | 1.37E-05 | 3.83E-06 |
| 465 | 1.27E-05 | 1.04E-05 | 465 | 1.33E-05 | 3.97E-06 |
| 470 | 1.33E-05 | 1.05E-05 | 470 | 1.28E-05 | 3.96E-06 |
| 475 | 1.40E-05 | 1.05E-05 | 475 | 1.23E-05 | 3.72E-06 |
| 480 | 1.46E-05 | 1.06E-05 | 480 | 1.20E-05 | 3.95E-06 |
| 485 | 1.58E-05 | 1.08E-05 | 485 | 1.18E-05 | 4.16E-06 |
| 490 | 1.71E-05 | 1.10E-05 | 490 | 1.18E-05 | 4.39E-06 |
| 495 | 1.83E-05 | 1.12E-05 | 495 | 1.19E-05 | 4.64E-06 |
| 500 | 1.96E-05 | 1.13E-05 | 500 | 1.24E-05 | 4.86E-06 |
| 505 | 2.12E-05 | 1.14E-05 | 505 | 1.27E-05 | 5.06E-06 |
| 510 | 2.26E-05 | 1.15E-05 | 510 | 1.26E-05 | 5.70E-06 |
| 515 | 2.38E-05 | 1.17E-05 | 515 | 1.19E-05 | 6.39E-06 |
| 520 | 2.52E-05 | 1.22E-05 | 520 | 1.12E-05 | 7.31E-06 |
| 525 | 2.69E-05 | 1.28E-05 | 525 | 1.02E-05 | 8.88E-06 |
| 530 | 2.86E-05 | 1.30E-05 | 530 | 9.22E-06 | 1.05E-05 |
| 535 | 3.21E-05 | 1.48E-05 | | | |